# Interconnecting bilayer networks

Xiu-Lian Xu, Yan-Qin Qu, Shan Guan, Yu-Mei Jiang, Da-Ren He[a]

*College of Physics Science and Technology, Yangzhou University, Yangzhou, 225002, P. R. China*

**Abstract** – A typical complex system should be described by a *supernetwork* or a *network of networks*, in which the networks are coupled to some other networks. As the first step to understanding the complex systems on such more systematic level, scientists studied interdependent multilayer networks. In this letter, we introduce a new kind of interdependent multilayer networks, i.e., interconnecting networks, for which the component networks are coupled each other by sharing some common nodes. Based on the empirical investigations, we revealed a common feature of such interconnecting networks, namely, the networks with smaller averaged topological differences of the interconnecting nodes tend to share more nodes. A very simple node sharing mechanism is proposed to analytically explain the observed feature of the interconnecting networks.

**Introduction.** - In the last decade, complex networks have been widely used in describing many kinds of complex systems, including ecological, social, biological and technological systems. Most of the previously discussed networks were limited to the isolated networks, which contain only one kind of nodes and interactions. However, Kurant and Thiran pointed out that "although the widely investigated networks are usually considered as distinct objects, they are often a part of larger complex systems, where a number of coexisting topologies interact and depend on each other" [1,2]. This means that a typical complex system should be described by a *supernetwork* [3] or a *network of networks* [4-6], in which the networks are coupled to each other. Such more systematic descriptions are becoming an attracting research area.

As the first step to understanding such complex *supernetworks*, scientists investigated interdependent multilayer networks recently [1,2,7,8]. For example, in Ref. [1], Kurant and Thiran introduced the layered model to describe transportation system, in which two network layers are used to represent the physical infrastructure and the traffic flows, respectively. With such layered idea, the underlying reason of the failure in estimating the real load of traffic network by the commonly used load estimator has been revealed. In this work, we introduce a new kind of multilayer networks, i.e., interconnecting networks. Different from the previously studied multilayer networks, in the interconnecting networks, the network layers are coupled together by sharing some nodes. The common nodes shared by the component networks are referred to as interconnecting nodes. It is common to observe such interconnection of network layers in real world complex systems. Without losing generality, for simplicity we will focus on the bilayer networks, which contain two network layers. For examples, some herbs (Chinese traditional medicines) may be simultaneously used as foods. Therefore, the network of Chinese herb prescription and the network of Chinese cooked dishes form one bilayer network (HP-CD bilayer) [9,10]. Some cities may have coach stations and railway stations. We can say that these cites (as the nodes) are shared by both the coach-traffic network and the train-traffic network, and the two traffic networks form one bilayer network (coach-train bilayer). Apparently, the interconnecting nodes may play very different roles in the two component networks. It is intuitionistic that the interconnecting networks with the interconnecting nodes playing more different roles in the component networks tend to have fewer interconnecting nodes and larger averaged topological differences of the interconnecting nodes in these component networks. This may be the most important general feature of such interconnecting networks. For example, the component networks in the coach-train bilayer (i.e., coach-traffic network and train-traffic network) are more similar than that in the HP-CD bilayer (i.e., network of Chinese herb prescription and the network of Chinese cooked dishes). Apparently, there are more interconnecting nodes in the coach-train bilayer than in the HP-CD bilayer. Similarly, it is rare that an important food (e.g., rice) is also an important herb, but it is quite common that a city having an important coach station can also have an important railway station. We will propose a very simple model to deduce the quantitative expression of this intuitionistic idea. Then the analytic results based on the proposed model will be compared with some empirical observations. Since the interconnecting networks are very general, the results of this study will be valuable for understanding the general properties of complex systems.

---

[a] E-mail: darendo10@yahoo.com.cn

**Data collection and empirical investigations.-**We collected eight bilayer networks (including sixteen single-networks) which are listed in Table 1. These bilayer networks cover wide categories, including food, biology, medicine, movie, and transportation. The details of data collection and network construction were presented in Ref. [11].

For quantitative discussions, we denote the nodes of the two layers by $V_1=\{i_1,i_2,\cdots,i_{M1}\}$ and $V_2=\{j_1,j_2,\cdots,j_{M2}\}$, respectively, and the two layers are referred to as upper layer and lower layer. By definition, we have $V_1 \cap V_2 \neq \varphi$. The node degree and the betweenness are used to describe the topology properties of the nodes. The node degree is defined as the number of edges connecting to the node, and the betweenness is defined as the number of the shortest paths passing through the node. For each interconnecting node, the topological difference is given by

$$u_x = \left| \frac{x_i}{\langle x_i \rangle} - \frac{x_j}{\langle x_j \rangle} \right|, \quad (1)$$

where $x_i$ and $x_j$ represent the node degree or betweenness of an interconnecting node in the upper layer and lower layer, respectively. $i$ and $j$ represent the indices of the same node in the respective layers. $\langle x_i \rangle$ and $\langle x_j \rangle$ denotes the averaged node degree ($x=k$) or betweenness ($x=b$) in all the nodes of the upper layer and lower layer, respectively. Accordingly, the averaged topological difference of the interconnecting nodes is given by

$$U_x = \frac{1}{m} \sum_{1}^{m} \left| \frac{x_i}{\langle x_i \rangle} - \frac{x_j}{\langle x_j \rangle} \right|, \quad (2)$$

where $m$ is the number of the interconnecting nodes. The normalized number of the interconnecting nodes is given by

$$n = \frac{2m}{M_i + M_j}, \quad (3)$$

where $M_i$ and $M_j$ are the total numbers of the nodes in the upper layer and lower layer, respectively.

For each of the eight bilayer networks, we empirically investigated the distributions of the node degree and betweenness. The details of the calculations are available in Ref. [11]. The results show that all the distributions can be fitted by the shifted power law (SPL) function, $p(x) \propto (x+\alpha)^{-\gamma}$, which interpolates between a power law function and an exponential function [10,12]. If $\alpha=0$, the shifted power law is reduced to a standard power law function. When $\alpha \rightarrow \infty$, the shifted power law is reduced to an exponential function. Therefore, the two parameters, $\alpha$ and $\gamma$, can be used to fully characterize the distributions of the node degree and the betweenness for each layer of the bilayer networks. We sort the real world bilayer networks by the ascend order of the $n$, i.e. $n_1 \leq n_2 \leq \cdots n_l \leq \cdots \leq n_8$. The fitted parameters of the distributions, the $U_x$ values, as well as the normalized number of the interconnecting nodes for all the eight bilayer networks are listed in Table 1 [11].

Table 1: Parameters of the eight bilayer networks. *M* and *e* are the node number and edge number of each network bilayer, respectively. *m* and *n* denote the number and the normalized number of the interconnecting nodes, respectively. $\alpha_k$, $\gamma_k$, $\alpha_b$, $\gamma_b$ denote the fitted SPL parameters of the degree distribution and betweenness distribution. $U_k$ and $U_b$ denote the averaged differences of the node degree and betweenness of the interconnecting nodes, respectively. $\gamma_{uk\text{-}ka}$ and $\gamma_{ub\text{-}ba}$ denote the fitted power exponents of the correlation between the $u_k$ and $k_a$, and between the $u_b$ and $b_a$, respectively. The details of deriving these parameters were presented in Ref. [11].

| Bilayer No. | Layer | $M$ | $e$ | $\alpha_k$ | $\gamma_k$ | $\alpha_b$ | $\gamma_b$ | $m$ | $n$ | $U_k$ | $U_b$ | $\gamma_{uk\text{-}ka}$ | $\gamma_{ub\text{-}ba}$ |
|---|---|---|---|---|---|---|---|---|---|---|---|---|---|
| 1 | HP | 1612 | 23035 | 0.15 | 3.22 | 0.065 | 2.046 | 43 | 0.039 | 2.69 | 3.76 | 0.72 | 0.24 |
|   | CD | 595 | 7876 | 0.0087 | 1.31 | 0.001 | 0.86 |   |   |   |   |   |   |
| 2 | BK | 4495 | 11183 | 0.008 | 2.11 | 0.035 | 1.601 | 169 | 0.045 | 1.20 | 3.41 | 1.09 | 1.18 |
|   | PK | 3028 | 5166 | 0.04 | 3.07 | 0.035 | 1.18 |   |   |   |   |   |   |
| 3 | MMA | 3219 | 44153 | 1 | 9.9 | 0.1 | 2.35 | 361 | 0.17 | 1.04 | 2.88 | 1.08 | 0.84 |
|   | HKMA | 1132 | 11455 | 0.475 | 5.9 | 0.1 | 2.25 |   |   |   |   |   |   |
| 4 | YPI | 3985 | 30677 | 0.016 | 2.10 | 0.005 | 1.440 | 445 | 0.20 | 0.97 | 1.32 | 0.97 | 0.91 |
|   | YM | 527 | 38285 | 10 | 81 | 0.21 | 2.89 |   |   |   |   |   |   |
| 5 | EPI | 2893 | 14009 | 0.01 | 1.79 | 0.004 | 1.328 | 623 | 0.34 | 1.03 | 1.29 | 0.90 | 0.49 |
|   | EM | 758 | 63035 | 10 | 100 | 0.55 | 5.48 |   |   |   |   |   |   |
| 6 | coach | 314 | 3220 | 0.37 | 4.1 | 0.15 | 2.72 | 100 | 0.48 | 0.72 | 1.07 | 0.89 | 0.87 |
|   | airplane | 100 | 838 | 1 | 5.5 | 0.05 | 1.39 |   |   |   |   |   |   |
| 7 | train | 251 | 6775 | 1 | 6.6 | 0.7 | 5.92 | 88 | 0.50 | 0.66 | 0.91 | 0.74 | 0.6 |
|   | airplane | 100 | 838 | 1 | 5.5 | 0.05 | 1.39 |   |   |   |   |   |   |
| 8 | coach | 314 | 3220 | 0.37 | 4.1 | 0.15 | 2.72 | 251 | 0.89 | 0.61 | 0.89 | 0.74 | 0.92 |
|   | train | 251 | 6775 | 1 | 6.6 | 0.7 | 5.92 |   |   |   |   |   |   |

From Table 1, we can see that with the increasing of *n*, the averaged topological difference $U_x$ (i.e., $U_k$ and $U_b$, for node degree and betweenness, respectively.) of the interconnecting nodes basically show monotonic decreasing. This qualitative observation indicates that the networks with smaller averaged topological differences of the interconnecting nodes tend to share more nodes. This is in qualitative agreement with the intuitionistic argument presented in the last section. In Fig. 1, the empirical correlation between the *n* and $U_x$ (triangle: $U_k$; cross: $U_b$) for all the eight bilayer networks are plotted. The analytic results are also given (solid lines), which will be explained in the next section. Such strong correlation between the *n* and $U_x$ may suggest some common features of the interconnecting nodes of each bilayer network.

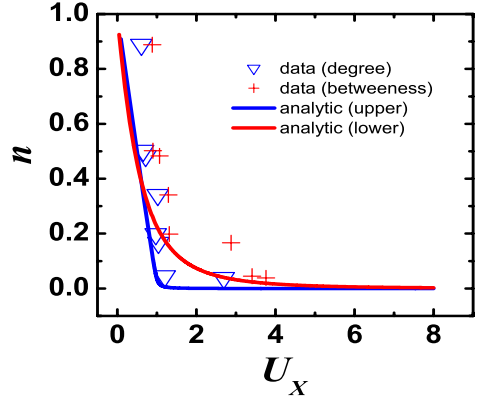

Fig. 1: (Color online) Analytic and empirical results of the relationship between $U_x$ and $n$.

**Model analysis and comparisons with empirical observations.-**We have tried several models for deducing the quantitative expressions of the empirical correlation between the $n$ and $U_x$ discussed in the last section. The results based on these models show very small differences (data not shown), suggesting that the general correlation between the $n$ and $U_x$ confers a strong tolerance to model details. In this situation, we prefer a model which presents the simplest explicit expression of the role differences of the interconnecting nodes among the component networks and the simplest mechanism for selecting $n$ and $U_x$ when the role difference of the interconnecting nodes varies.

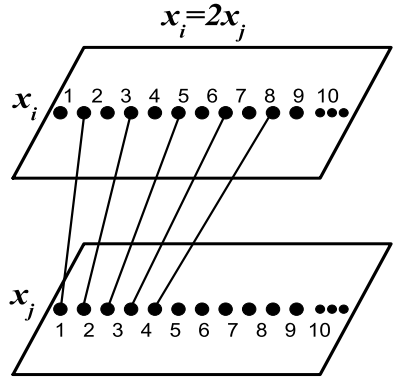

Fig. 2: A schematic graph showing the main idea of the model.

In the model, we adopt the following assumptions. Firstly, the upper and lower layers have exactly the same node numbers ($M_i=M_j=M$) and the same distributions of node degree and betweenness, therefore, $<x_i>=<x_j>=<x>$. Secondly, the $x$ values of all the interconnecting nodes in the upper layer is larger than that in the lower layer, therefore, $u_x=(x_i-x_j)/<x>$. Thirdly, the principle, by which the node $j$ in the lower layer selects and merges to the node $i$ in the upper layer so as to form an interconnecting node, is given by $x_i=\xi(l)x_j$ where $\xi(l)$ is a constant depending on the index $l$ of the bilayer systems. This model can be understood from the schematic graph shown in Fig. 2 where the nodes in the upper and lower layers are represented by solid circles and labeled in ascend order of $x$ values. For demonstration, the value of the $\xi(l)$ is chosen as 2 in Fig.2. The lines connecting the nodes from both layers represent that the two nodes are one interconnecting node. It is easy to image that with a larger coefficient $\xi(l)$, the slope of the line decreases more quickly with the increasing of the $x$. Consequently, fewer nodes can be shared by the two layers, and simultaneously, the averaged topological differences of the interconnecting nodes are larger, which is consistent with the observed correlation between the $n$ and $U_x$, namely, the networks with larger averaged topological differences of the interconnecting nodes tend to share more nodes. Apparently, $\xi(l)$ is the simplest explicit expression of the role difference of the interconnecting nodes among the component networks and the model expresses the simplest mechanism for selecting $n$ and $U_x$ when the role difference of the interconnecting nodes varies. We will show that the empirically observed relation between the $n$ and $U_x$ can be derived analytically based on the above model.

According to the definition of the distributions of node degree or betweenness, the total number of the interconnecting nodes can be calculated by

$$m = \sum_{\xi(l)x_{\min}}^{x_{\max}} Mp(x), \tag{4}$$

where $x_{min}$ and $x_{max}$ are the minimal and maximal values of the node degree or betweenness, respectively. Substitute Eq.3 into Eq.4, we can get the normalized number of the interconnecting nodes by

$$n(\xi(l)) = \frac{m}{M} = \sum_{\xi(l)x_{\min}}^{x_{\max}} p(x_i) \cong (\int_{\xi(l)x_{\min}}^{x_{\max}} c(x_i/x_{\max}+\alpha)^{-\gamma-1}dx_i)/\xi(l) = \frac{E(\xi(l))}{\xi(l)B}, \tag{5}$$

where $A=x_{min}/x_{max}$, $B=(A+\alpha)^{-\gamma}-(1+\alpha)^{-\gamma}$, $E(\xi(l)) = [\xi(l)A+\alpha]^{-\gamma} - (1+\alpha)^{-\gamma}$, and $c=\gamma/(x_{max}B)$. Similarly, the averaged topological difference of the interconnecting nodes $U_x$ can also be obtained as a function of $\xi(l)$ by

$$U_x = \frac{\sum_{\xi(l)x_{\min}}^{x_{\max}} Mp(x_i)\frac{x_i-x_j}{\langle x \rangle}}{m} \cong \frac{cx_{\max}[\xi(l)-1]}{\langle x \rangle n\xi^2(l)} \int_{\xi(l)x_{\min}}^{x_{\max}} [(x_i/x_{\max}+\alpha)^{-\gamma} - \alpha(x_i/x_{\max}+\alpha)^{-\gamma-1}]dx_i. \tag{6}$$

Considering that $<x>$ can be calculated by

$$\langle x \rangle \cong \int_{x_{\min}}^{x_{\max}} xp(x)dx = \frac{\gamma x_{\max}D}{(\gamma-1)B} - \alpha x_{\max}, \tag{7}$$

with $D=(A+\alpha)^{-\gamma+1}-(1+\alpha)^{-\gamma+1}$, we get

$$U_x(\xi(l)) = \frac{[\xi(l)-1]B[\gamma F(\xi(l))-\alpha(\gamma-1)E(\xi(l))]}{\xi(l)E(\xi(l))[\gamma D-\alpha(\gamma-1)B]}, \tag{8}$$

where $F(\xi(l)) = [\xi(l)A+\alpha]^{1-\gamma} - (1+\alpha)^{1-\gamma}$. Combining (5) and (8), one can obtain the quantitative relationship between the $n$ and $U_x$, which can then be compared with the empirical results.

In Fig. 1, the upper solid line is the analytical result (by (5) and (8)) with the parameters $\alpha=0.001$, $\gamma=0.87$, $x_{\min}=1$, and $x_{\max}=500$. The lower line is the results with the parameters $\alpha=10$, $\gamma=100$, $x_{\min}=1$, and $x_{\max}=5000$. The values of the parameters $\alpha$ and $\gamma$ in the upper line and lower line correspond to the minimal values and maximal values observed empirically. The value of the $x_{\max}$ does not affect the results when it is large enough. One can see that, even with such a simple model, the analytic conclusion show rather good agreement with the empirical results, suggesting that the above model captures the essential features of the interconnecting bilayer networks. Meanwhile, these results also suggest that the correlation between the $n$ and the $U_x$ is the common property of the discussed interconnecting networks, namely, the networks with larger averaged topological differences of the interconnecting nodes tend to share more nodes.

**Summary and discussions.-** In summary, we introduced a new kind of multilayer networks, i.e., interconnecting networks, in which the component network layers share some common nodes. We emphasize that such interconnecting networks are conceptually different from the multilayer networks discussed in literatures [1,2,7,8]. What discussed in these previous works are some kinds of functional interdependences between the network layers, while the current study discusses the merging of a part of the nodes, which can be considered as a special interdependence but is very ubiquitous. Based on the empirical investigations of eight bilayer networks, we revealed, by a very simple model, a common feature of such layered networks, namely, the networks with larger averaged topological differences of the interconnecting nodes tend to share more nodes. This conclusion can be readily extended to general multilayer networks although we discussed only bilayer networks in this work. In addition, although the feature revealed in this work is about the relationship of two topological properties, $n$ and $U_x$, the model clearly shows that, in a statistical sense, the topological properties depend on network functions. It is the role difference of the interconnecting nodes that determines both the $n$ and $U_x$. In the current work, we used the node degree and betweenness to represent the topological properties of the interconnecting nodes. We also tried the averaged nearest-neighbor degree and the node strength. The empirical results of the $n$-$U_x$ relationships are also in reasonable agreement with the analytic results (data not shown).

Concerning the kernel assumption of the model, some questions may arise. For example, some people may think that the assumption of $x_i=\xi(l)x_j$ in the last section is too simple to be accepted. Can we provide some, even indirect, empirical supports to this assumption?

A direct empirical verification on the relation $x_i=\xi(l)x_j$ is difficult due to the large fluctuations as we have already tried. Here we provide an indirect empirical verification to the above model assumption. As an averaged consideration that may depress the fluctuations, we compared the $u_x$ and the $x_a=(x_i+x_j)/2$ for all the interconnecting nodes of each bilayer network, which describe the topological difference of the interconnecting nodes and the average of the topological properties of each node in the two layers, respectively. Fig. 3 shows the results for one representative bilayer network, i.e., YPI-YM bilayer [11]. This bilayer is composed of two network layers: protein interaction network in yeast (YPI) (upper layer) [13,14] and metabolic network in yeast (YM) (lower layer) [15]. In the upper layer, nodes are defined as proteins and links are defined as physical interactions between proteins. In the lower layer, nodes are defined as enzymes (a kind of proteins) and links are defined as common biochemical compounds shared by two enzymatic reactions. The interconnecting nodes are defined as the proteins shared by both protein interaction network layer and metabolic network layer. The results for all the other bilayer networks were given in Ref. [11]. Empirically, we find that all the $u_x$-$x_a$ data can be fitted by the power law function with the scaling exponents around 1.0 as shown by the two columns in the most right of Table 1. In order to further eliminate fluctuations, cumulative relations are calculated and shown in Fig. 3. The empirical results suggest a relation, $\lg u_x = (1\pm\delta(l))\lg x_a + \nu(l)$, where $\delta(l)$ is a small quantity. By ignoring $\delta(l)$, we get $u_x \cong \eta(l) x_a$ where $\eta(l)=10^{\nu(l)}$ is a proportional coefficient depending on index $l$ of the bilayer system. Accordingly, we have $(x_i-x_j)/<x> = \eta(l)x_a = \eta(l)(x_i+x_j)/2$, which immediately lead to $x_i=\xi(l)x_j$ with $\xi(l)=[1+\eta(l)<x>/2]/[1-\eta(l)<x>/2]>1$ depending on the index $l$ of the bilayer system.

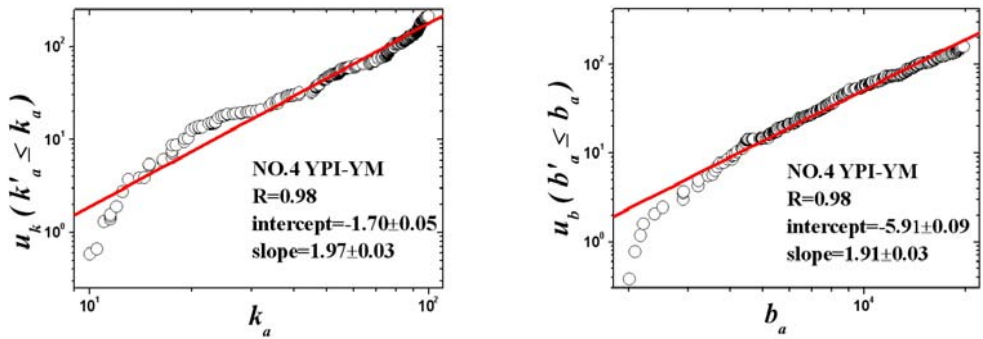

Fig. 3: $u_k$-$k_a$ (a) and $u_b$-$b_a$ (b) relations obtained empirically for a representative interconnecting network, YPI-YM bilayer. The solid lines are the results of the least square fitting to all the data.

The feature revealed in the current work for this new kind of multilayer networks will be valuable for understanding the complex systems in a more systematic level. It is reasonable to expect that the interconnection of some network layers may change the important properties or show some new properties of some real world complex systems, which may in turn induce new understanding and applications. More quantitative investigations to such interconnecting networks are needed in the future work.

***

We acknowledge Prof. Z.-R. Di (Beijing Normal University) and Prof. T. Zhou (Electronic Scientific and Technological University of China) for very helpful discussion. This work is supported by the Chinese National Natural Science Foundation under grant numbers 10635040 and 70671089.